\documentclass[showpacs,amssymb,eqsecnum,aps]{revtex4}
\usepackage{color}
\newcommand{\be}{\begin{equation}}
\newcommand{\ee}{\end{equation}}

\newcommand{\nnbb}{\nonumber\\}

\newcommand{\tr}{{\rm Tr\,}}

\newcommand{\sun}{${\rm SU}(N)\;$}

\newcommand{\tpte}{\theta\, p\, T}
\begin{document}

\title{Transport equation and hard thermal loops
in noncommutative Yang-Mills theory}

\author{F. T. Brandt$^a$, Ashok Das$^b$, J. Frenkel$^a$,
D. G. C. McKeon$^{c}$ and J. C. Taylor$^{d}$}
\affiliation{$^a$ Instituto de F\'{\i}sica,
Universidade de S\~ao Paulo,
S\~ao Paulo, SP 05315-970, BRAZIL}
\affiliation{$^b$Department of Physics and Astronomy,
University of Rochester,
Rochester, NY 14627-0171, USA}
\affiliation{$^c$ Department of Applied Mathematics, The University of
Western Ontario, London, ON  N6A5B7, CANADA}
\affiliation{$^d$ Department of Applied Mathematics
and Theoretical Physics, University of Cambridge,
Cambridge, UK \\ }

\date{\today \\ }

\begin{abstract}
We show that the high temperature limit of the noncommutative thermal
Yang-Mills theory can be directly obtained from the Boltzmann transport
equation of classical particles. As an illustration of the simplicity
of the Boltzmann method, we evaluate the two and the three-point gluon
functions in the noncommutative $U(N)$  theory at high
temperatures $T$. These amplitudes are gauge  invariant and satisfy
simple Ward identities. Using the constraint satisfied at order $T^2$
by the covariantly conserved current, we construct the hard thermal
loop effective action of the noncommutative theory.
\end{abstract}


\maketitle

\section{Introduction}

In recent years, there has been a lot of interest in the study of the  
dynamics of hot plasma in non-Abelian gauge theories. It is known now 
that the $n$-point gluon functions in such a theory, when 
evaluated at temperatures which are large compared with all 
momenta $p$ of the external gauge fields, are gauge invariant and 
have a leading behavior proportional to  $T^2$. Here, $T$ represents
the temperature of the plasma. In order to account for all the leading
order contributions consistently as well as to obtain a meaningful
gauge invariant result for physical quantities, it is  
necessary to perform a resummation of hard thermal loop 
contributions \cite{Braaten:1990it}.
The hard thermal loops are defined by the relation
\begin{equation}
p\ll k \sim T\label{hard}
\end{equation}
where $p$ represents a characteristic external momentum while $k$
denotes the internal loop momentum. In conventional QCD, such a
procedure,  although  very insightful, is technically quite involved.
However, it has also been noted that it is possible to use a classical
transport equation in a  more direct and transparent way, in order to
derive  the effective action which includes all the contributions
associated  with the hard thermal loops
\cite{Heinz:1985yq,Blaizot:1994be,kelly:1994ig,Litim:2001db}. 
In this approach, which has been
successfully applied to a variety of  gauge theories, one pictures the
constituents of the plasma as classical charged particles interacting
in a  self-consistent manner. The main reason why such a classical
description works in yielding a quantum effective action is that, for
soft gauge fields, the occupation number per unit mode, in a hot
plasma, is quite high due to the Bose-Einstein enhancement.

In more recent years, developments in string theory have renewed
interest in noncommutative field theories \cite{Seiberg:1999vs,Fischler:2000fv,Arcioni:1999hw,Landsteiner:2000bw,Szabo:2001kg,Douglas:2001ba}.
These are theories
defined on a noncommutative manifold satisfying
\begin{equation}
[x^{\mu},x^{\nu}] = i \theta^{\mu\nu}
\end{equation}
where $\theta^{\mu\nu}$ is assumed to be a constant anti-symmetric
tensor. Furthermore, to avoid problems with unitarity, one also
assumes that $\theta^{0\mu} = 0$, namely, only the spatial coordinates
do not commute. Note that, by definition, $\theta^{\mu\nu}$ has the
dimensions of the square of a length. Quantum field theories defined
on such  a manifold --
noncommutative field theories -- exhibit some very interesting features
and there are continued attempts at understanding better the structures
of such theories, in particular, those of noncommutative gauge theories.

In the paper \cite{Brandt:2002aa}, the structure of the two- and three-gluon
amplitudes, incorporating the hard thermal loop contributions, were
calculated in the pure noncommutative $U(N)$ Yang-Mills theory at high
temperature and the results were quite interesting. We note that while
(\ref{hard}) defines the hard thermal loops, in the presence of
independent dimensional parameters, such as $\theta^{\mu\nu}$, we can
also have regimes satisfying $\theta < {1\over T^{2}}$ or $\theta >
{1\over T^{2}}$, in addition to Eq. (\ref{hard}). Here, $\theta$ can be
thought of as the magnitude of $\theta^{\mu\nu}$ (or, in more physical
terms as the largest or the smallest eigenvalues of $\theta^{\mu\nu}$
for the first and the second inequalities respectively). In
the presence of a characteristic momentum scale, $p$, one can also have
regimes satisfying $\theta pT < 1$ or $\theta p T > 1$. In fact, it
was shown in \cite{Brandt:2002aa} that the calculation of hard thermal loops simplifies
enormously in the two limits $\theta p T \ll 1$ and
$\theta p T \gg  1$. We note that, in the presence of a dimensional
parameter, such as $\theta$, the hard thermal loop condition,
Eq. (\ref{hard}), can be written as
\begin{equation}
 \theta p^{2} \ll \theta p T \ll \theta T^{2} \label{weak}
\end{equation}
The actual hard thermal loop calculations \cite{Brandt:2002aa} show that, the
leading  $T^2$ contributions from the $U (1)$ sector of the
$U(N)$ theory become suppressed by powers of $\theta p T$ when $\theta
p T\ll 1$, while, in the limit $\theta p T\gg 1$, the amplitudes are
all  proportional to $N$ as would be the case for a large $N$ theory.

It is worth noting that a similar behavior (at zero temperature) is
present in the case of a noncommutative supersymmetric $U(N)$ gauge theory
\cite{Chu:2001fe}.  Namely, it was shown there that the $U(1)$ sector of the
theory behaves differently for $p<\theta^{-{1\over 2}}$ and
$p>\theta^{-{1\over 2}}$ such that in the low energy limit, the
noncommutative $U(N)$ theory behaves like an ordinary $SU(N)$ theory
(We prefer ``ordinary'' to ``commutative'' for obvious reasons.). As is
the case in \cite{Chu:2001fe}, here, too, we will like to emphasize that the
behavior  of the theory in the hard thermal loop approximation should
not be taken to imply that temperature somehow breaks the $U(N)$
symmetry. In fact, in connection with the work of \cite{Chu:2001fe}, gauge
invariant completions of the action involving open Wilson lines have
already been proposed in \cite{VanRaamsdonk:2001jd,Armoni:2001uw}. 
Physically, the results of
\cite{Brandt:2002aa}, 
in both the limits, may be understood as follows. The limit, $\theta p
T\ll 1$, can clearly be thought of as corresponding to weak
non-commutativity, in
which case, the self couplings of the $U(1)$ sector is negligible
compared to that of the $SU(N)$ sector. Therefore, the theory, to the leading
order, behaves like an ordinary $SU(N)$ theory with the effects of
non-commutativity leading to small corrections. In the limit
$\theta p T\gg 1$, on the other hand, we can think of the
non-commutativity as strong and it is well known \cite{Minwalla:1999px} that in
the  limit of
maximal non-commutativity, a noncommutative theory has a stringy
behavior. (Intuitively, this is seen as follows. If the noncommutative
theories are supposed to describe dipoles \cite{Sheikh-Jabbari:1999vm,Bigatti:1999iz}, then, for
distances much
much smaller than the dipole length, ${1\over T}\ll \theta p$, the
string nature would become manifest.) In particular, in the case of a
noncommutative gauge 
theory, this would correspond to having dominant planar diagrams. (The
non-planar diagrams oscillate rapidly at high temperature and yield
only subleading contributions.)
        
It is an interesting question to ask whether a classical transport
equation can be written for such theories which can determine the
effective action in the hard thermal loop approximation in a simple
manner, as is the case for ``ordinary'' theories. Of course, as
we have mentioned earlier, there are two limits, $\theta p T\ll 1$ and
$\theta p T\gg 1$, in which one can analyze the effective action. Since the
region $\theta p T\ll 1$ corresponds to a weaker non-commutativity, 
in this paper, we restrict ourselves to the region  
where $\tpte\gg 1$ and which genuinely exhibits noncommutative hard 
thermal loop  effects. We will present a classical transport equation
for the noncommutative $U(N)$ gauge theory which leads, in a simple
way, to the leading order terms of the two and three gluon amplitudes
in the  hard thermal loop approximation, calculated in \cite{Brandt:2002aa}. The color
current, defined as
\be\label{eqJ}\label{4a}
J_\mu^C(x,A) = {\delta \Gamma\left[{\bf A}\right]\over \delta A^C_\mu(x)}
\ee
where $\Gamma\left[{\bf A}\right]$ represents the effective action in the
hard thermal loop approximation, can be derived from the classical
transport equation. Here, $A_{\mu}^{C}$ is the gluon field and
$J_{\mu}^{C}$ consists of an infinite set of graphs. The above equation
can be integrated and leads to the gauge invariant
effective action  which incorporates all the hard thermal 
loop contributions.

The paper is organized as follows. In section {\bf II}, we derive the
transport equation for the noncommutative $U(1)$ theory in two
alternate ways. The first is through the conventional use of the
dynamical equations for a relativistic particle in a noncommutative
manifold. The second is through the use of the conservation equations
for the current as well as the stress tensor. Both these methods
yield the same transport equation which reduces, in the $\theta\rightarrow 0$
limit, to the conventional transport equation for QED. In the limit
$\theta p T\gg 1$, we show how this transport equation can be used to
derive, in a simple manner, the two and the three point functions for
the $U(1)$ gauge boson, which agrees with the diagrammatic
calculations in \cite{Brandt:2002aa}. 
The derivation of the transport equation for the $U(N)$ case is rather
involved for various technical reasons. We do not fully understand a
first principle derivation in this case. In section
{\bf III}, we discuss some of the challenges that one faces in the
$U(N)$ case and with guidance from the $U(1)$ case, propose a
transport equation  for the noncommutative
$U(N)$ theory. This equation reduces to that
for the conventional $SU(N)$ theory in the limit $\theta\rightarrow
0$. However, in the limit $\theta p T\gg 1$, it leads to genuine
noncommutative hard thermal loop effects (which have a stringy
character). We evaluate the two and three point gluon functions from
the transport equation and these agree with the explicit diagrammatic
calculations in \cite{Brandt:2002aa}. 
In section {\bf IV}, we derive from the transport equation
an all order expression for the
color current [Eq. (\ref{35})], which is manifestly covariantly conserved. 
With the help of this result, we obtain
a gauge invariant form for the effective action [Eq. (\ref{55})]
which includes all the hard thermal loop effects in noncommutative 
Yang-Mills theory.

\section{Transport equation for noncommutative $U(1)$ theory}

In this section, we consider the noncommutative $U (1)$ gauge
theory without any matter fields. As is well known, such a theory is
self-interacting much like the conventional Yang-Mills theories. In
this case, the field strength is defined to be
\be\label{eqF}
F_{\mu\nu}=\partial_\mu\,A_\nu - \partial_\nu\,A_\mu - 
i\,g\,[A_\mu,A_\nu]_{{\rm M}}
\ee
where
\begin{equation}
[A_{\mu},A_{\nu}]_{\rm M} = A_{\mu}\star A_{\nu} - A_{\nu}\star
A_{\mu}\label{moyal}
\end{equation}
defines the Moyal commutator and the star product that introduces
self-interactions into the system is given by 
\be
A_\mu(x)\star A_\nu(x) = \left[
{\rm e}^{\frac i 2 \, \theta^{\alpha\beta}\partial_\alpha^{(\xi)}
                                          \partial_\beta^{(\eta)}}\,
A_\mu(x+\xi) \, A_\nu(x+\eta)
\right]_{\xi=\eta=0}
\ee

The Euler-Lagrange equation, in this case, is obtained to be
\be\label{EL}
\partial^\mu\,F_{\mu\nu} -i\,g\, [A^\mu,F_{\mu\nu}]_{{\rm M}}\equiv
{\cal D}^\mu\,F_{\mu\nu} = J_\nu.
\ee
where $J_{\mu}(x)$ represents the source for the gauge fields and
${\cal D}_\mu$ denotes  the  Moyal covariant derivative. It follows
from Eq. (\ref{EL}) that the current $J_\mu$ is  covariantly
conserved.  This property ensures that the effective
action, which leads to the current through a functional derivation,
namely, 
\be\label{4}
J^\mu(x,A) = {\delta\Gamma\left[{ A}\right]\over\delta A_\mu(x)},
\ee
is gauge invariant. This follows since, under an infinitesimal
noncommutative $U(1)$ gauge transformation, 
\be\label{5}
{\delta\Gamma\left[{ A}\right]\over\delta \omega (x)} = \int dy\,{\delta
A_\mu(y)\over\delta\omega (x)} {\delta\Gamma\left[{ A}\right]\over\delta A_\mu(y)} = 
 {\cal D}_\mu\,J^\mu(x) = 0.
\ee
Here, $\omega (x)$ represents the infinitesimal parameter of gauge
transformation. Let us note that the covariant conservation of
the current is a consequence (as well as a reflection) of the fact
that, in the noncommutative $U(1)$ theory, the current transforms in
the adjoint representation under a star gauge transformation,
\begin{equation}
J_{\mu}(x) \rightarrow U(x)\star J_{\mu}(x)\star
U^{-1}(x)\label{current}
\end{equation}
This has to be contrasted with the behavior of the current in an
``ordinary'' $U(1)$ theory.

There are several ways one can derive the transport equation for the
noncommutative $U(1)$ theory. Following the conventional method, we
define the current, $J^\mu(x)$, in terms of the statistical distribution 
functions, $f(x,k)$, as
\be\label{6}
J_\mu(x) = g\, \sum_{{\rm }} \int dK\, k_\mu\, f(x,k),
\ee     
where the sum is over contributions from all particle species and
helicities. Let us note that, unlike in conventional $U(1)$ theory, in
the noncommutative case, charge neutral particles belong to the
adjoint representation of the group and contribute to the current as
well. In fact, in the present case, where we are investigating a pure
$U(1)$ gauge theory (noncommutative), there is no charged matter
present and the gauge field is charge neutral so that the classical
particle that we are considering also is charge neutral (nonetheless
has a nontrivial current). Furthermore, the integration  measure for
momentum is defined to  be
\be\label{7}
dK \equiv {d^4 k\over (2\pi)^3} 2\, \theta(k_0)\,\delta(k^2-m^2)
\ee
which guarantees positivity of the energy as well as the on-shell
evolution  of the thermal excitations. Let us note next from
Eqs. (\ref{current}) and the definition of the current in (\ref{6})
that, in this case, the distribution function $f(x,k)$
transforms covariantly under a star gauge transformation much like the
current, namely
\be\label{8}
 f(x,k) \rightarrow  U(x) \star f(x,k)\star U^{-1}(x). 
\ee
Once again, this behavior has to be contrasted with the conventional
case and we note that when $\theta^{\mu\nu}\rightarrow 0$ and the star
product reduces to ordinary products, we recover the conventional
behavior for the distribution function.

In the case of a collisionless plasma, the distribution function
$f(x,k)$  may be determined from a classical collisionless transport
equation, which can be derived in a simple manner as follows. Since
$f(x,k)$ transforms in the adjoint representation, Eq. (\ref{8}), much
like the current, it follows that the distribution function must also
be covariantly constant along the trajectory of the particle, namely,
\be\label{9}
{\cal D}_\tau\, f(x,k) = {df\over d\tau} - ig {dx^{\mu}\over d\tau}
[A_{\mu},f]_{\rm M} = 0,
\ee
Here $\tau$ is the proper time which parameterizes the trajectory and
${\cal D}_\tau$ is the Moyal covariant derivative along the trajectory
of the  particle. Furthermore, using the equations of motion
\be\label{10}
m{dx^\mu\over d\tau} = k^\mu;\qquad
m {d k^\mu\over d\tau} = g\, F^{\mu\nu}\, k_\nu
\ee
which hold unchanged in the noncommutative theory,  
Eq. (\ref{9}) leads to the transport equation,
\be\label{11}
m{df\over d\tau} = \left(k\cdot\partial\right) \, f - g\,k_\mu\,
F^{\mu\nu}\star{\partial f\over\partial k^\nu} = i\, g\,
\left[k\cdot A, f\right]_{{\rm M}},
\ee
This may also be written in the simpler form,
\be\label{12}
\left(k\cdot {\cal D}\right) f(x,k) = g\, k_\mu\, F^{\mu\nu}\star{\partial f
\over \partial k^\nu}
\ee
This equation, which generalizes the Boltzmann equation for 
the ``ordinary''  $U (1)$ gauge theory, has the correct covariance 
properties under star gauge transformations, which characterize the
noncommutative theory.

The transport equation can also be derived alternatively without
using directly the equations of motion of the particle. For example,
we have already noted that the current can be written in terms of the
statistical distribution function. Similarly, generalizing the
conventional ideas, we also observe that the stress tensor for
the theory can also be described in terms of the function $f(x,k)$ as 
\begin{eqnarray}
J_{\mu} (x) & = & g \sum \int dK\ k_{\mu} f(x,k)\nonumber\\
T_{\mu\nu} (x) & = & \sum \int dK\ k_{\mu}k_{\nu} f(x,k)\label{stress}
\end{eqnarray}
It can be easily checked that requiring the conservation equations for
the  current and the stress tensor to hold, namely,
\begin{eqnarray}
{\cal D}_{\mu} J^{\mu} & = & 0\\
{\cal D}_{\mu} T^{\mu\nu} & = &  -F^{\mu\nu}\star
J_{\mu}\label{conservation}
\end{eqnarray}
also leads to the transport equation (\ref{12}). We note that, unlike
in the conventional $U(1)$ theory, here the stress tensor transforms 
in the adjoint representation
which is the reason for the (Moyal) covariant derivative in
\hbox{Eq. (\ref{conservation})}. It is worth pointing out that, in the limit
$\theta^{\mu\nu}\rightarrow 0$, we recognize Eq. (\ref{12}) to
correspond to the transport equation for conventional QED. In the
present case of a pure gauge theory, the self-interactions 
of the gauge bosons go to zero in this limit so that we do not expect hard
thermal  loop corrections in the absence of charged particles in
\hbox{Eq. (\ref{stress})}.
That this follows from the transport equation as well, is
seen easily from the fact that 
since we are considering a charge neutral particle (which belongs to
the adjoint representation of the star gauge group), in the vanishing
$\theta$ limit, the current associated with it also vanishes and the
transport equation, (\ref{12}) becomes trivial.

As an example of the convenience of this method in the evaluation 
of the hard thermal loop contributions, we will derive next the leading order  
contributions to the two- and three-point functions in the hard
thermal loop approximation, using the transport equation. 
Let us expand the distribution function in a power series in $g$
\be\label{13}
f = f^{(0)}+ g\, f^{(1)} + g^2\,f^{(2)} + \cdots,
\ee
where $f^{(0)}$ is the equilibrium Bose-Einstein 
distribution function
\be\label{14}
f^{(0)}(k_0) = {1\over{\rm e}^{|k_0|/T} -1}.
\ee
On the other hand, $f^{(1)}$ and $f^{(2)}$ represent the leading 
order corrections to this function, which can be
determined order by order from the transport equation, Eq. (\ref{12}),
to be
\be\label{15}
f^{(1)}(x,k) = {1\over k\cdot\partial}
\left(k\cdot\partial\, A^\nu - \partial^\nu\, k\cdot A\right)
{\partial f^{(0)}\over \partial k^\nu}
\ee
and
\begin{eqnarray}\label{15b}
f^{(2)}(x,k) & = & {i\over k\cdot \partial} \left([k\cdot A,
f^{(1)}]_{\rm M} - [k\cdot A, A^{\nu}]_{\rm M} {\partial
f^{(0)}\over \partial k^{\nu}} - i (k\cdot \partial A^{\nu} -
\partial^{\nu}k\cdot A)\star {\partial f^{(1)}\over \partial
k^{\nu}}\right)\nonumber\\
 & = &  {i\over k\cdot\partial}\left(
\left[k\cdot A, f^{(1)}\right]_{{\rm M}} - 
\left[k\cdot A, A^\nu\right]_{{\rm M}}
{\partial f^{(0)}\over \partial k^\nu} + \cdots
\right),
\end{eqnarray}
where $\cdots$ represent the third term which we have neglected since
it leads to a subleading contribution at high
temperature. 

Substituting these into the definition of the current in
(\ref{6}), we obtain the current as a series in powers of the gauge
field. Transforming to the momentum space and using the definition in
Eq. (\ref{4}), we obtain the two point function as
\be\label{16}
\Pi_{\mu\nu}(p) = \left.{\delta^{2}\Gamma\left[{ A}\right]\over \delta A^{\mu}(p)\delta
A^{\nu}(-p)}\right| = \left.{\delta J_{\mu}(p)\over \delta
A^{\nu}(-p)}\right| = -{2\,g^2\over (2\pi)^3}\int {d^3 k\over |\vec k|}
{1\over {\rm e}^{|\vec k|/T} -1}\left(
{p^2 k_\mu k_\nu\over (k\cdot p)^2} -
{p_\mu k_\nu + p_\nu k_\mu\over k\cdot p} + \eta_{\mu\nu}
\right),
\ee
where $k_\mu=|\vec k|(1,\hat k)\equiv |\vec k|\,\hat k_\mu$ denotes the
on-shell 4-momenta of the thermal particle in the plasma. Here, the
restriction on the functional derivatives stands for setting all the
fields to zero after differentiation. We note that
the  expression inside the parenthesis in (\ref{16}) is a homogeneous
function  of $k_\mu$ of degree zero, so that it is actually
independent of the magnitude $|\vec{k}|$. This  fact allows us to do 
the integration over $|\vec{k}|$ in (\ref{16}) leaving us
with an angular integral, which at high temperatures has the form
\be\label{17}
\Pi_{\mu\nu}(p) = - {g^2\, T^2\over 24\,\pi}\int {d\Omega}
\left({p^2 \hat k_\mu \hat k_\nu\over (\hat k\cdot p)^2} -
{p_\mu \hat k_\nu + p_\nu \hat k_\mu\over \hat k\cdot p} + \eta_{\mu\nu}
\right)\equiv
-{g^2\, T^2\over 24\,\pi}\int d\Omega\, G_{\mu\nu}(p,\hat k),
\ee
where $d\Omega$ denotes integration over all angular directions of the
unit vector $\hat k$ and we have, for simplicity, defined
\begin{equation}
G_{\mu\nu} (p,\hat{k}) =  \left({p^2 \hat k_\mu \hat k_\nu\over (\hat k\cdot p)^2} -
{p_\mu \hat k_\nu + p_\nu \hat k_\mu\over \hat k\cdot p} + \eta_{\mu\nu}
\right)\label{G}
\end{equation}
As expected, the two point function is independent of the parameter of
non-commutativity, $\theta^{\mu\nu}$, and is manifestly
transverse. Furthermore,  this result agrees
completely with the results from the explicit perturbative
calculations in \cite{Brandt:2002aa}  for the case $\theta p T\gg 1$.

Proceeding along similar lines, we can also obtain the three-point 
function which, to leading order, has the form
\be\label{18a}
\Gamma^\theta_{\mu\nu\lambda} =  \left.{\delta^{2}J_{\mu}(p_{1})\over
\delta A_{\nu}(p_{2})\delta A_{\lambda}(p_{3}=-(p_{1}+p_{2}))}\right| = \sin
\left({p_{1 \alpha}\,\theta^{\alpha\beta} p_{2 \beta}\over 2}\right)  
\Gamma_{\mu\nu\lambda}(p_1,p_2,p_3)
\ee
where $\theta^{\alpha\beta}$ is the parameter of non-commutativity and
\be\label{18}
\Gamma_{\mu\nu\lambda}(p_1,p_2,p_3) = -{i\,g^3\, T^2\over 24\pi}
\int d\Omega{1\over\hat k\cdot p_3}\left[
G_{\mu\nu}(p_2,\hat k)\hat k_\lambda  + \hat k_\mu \hat k_\nu 
H_\lambda (p_2,p_3,\hat k) - (p_1\leftrightarrow p_2)\right].
\ee
$G_{\mu\nu}$ was already given in Eq. (\ref{G}) and we have defined
the vector
\be\label{19}
H_\lambda(p_2,p_3,\hat k) = {1\over\hat k\cdot p_2}\left(
{p_2\cdot p_3\over\hat k\cdot p_3}\hat k_\lambda - {p_2}_\lambda\right)
\ee
As expected, the parameters  $\theta^{\mu\nu}$ are explicitly present
in the three point function reflecting the star product in the interaction terms. 
With the help of some simple algebraic identities, 
the expression (\ref{18}) can also be written
in a manifestly Bose symmetric form as follows
\begin{eqnarray}\label{30p}
\Gamma_{\mu\nu\lambda}(p_1,p_2,p_3) = {i\, g^3\, T^2\over 24\, \pi}
\int d\Omega\left[{1\over 2}{p_1^2\over (\hat k \cdot p_1)^2}
\left({1\over \hat k \cdot p_2} - {1\over \hat k \cdot p_3}\right)
  \hat k _\mu\,\hat k _\nu\,\hat k _\lambda \right. \nnbb \left.
+ \left({{p_1}_\lambda\over \hat k \cdot p_1} - {{p_2}_\lambda\over \hat k \cdot
    p_2}\right) {\hat k _\mu\,\hat k _\nu\over \hat k \cdot p_3}
+{\rm two\; cyclic\; permutations}\right]
\end{eqnarray}
The above form  for the three-point function is in complete  
agreement with the hard thermal loop results \cite{Brandt:2002aa} obtained by
standard  Feynman
diagrammatic calculations in the region $\tpte\gg 1$.
The expression (\ref{30p}) is the same as the one which occurs in the
momentum dependent part of the corresponding quantity in commutative
QCD (where the ``$\sin$'' in (\ref{18a}) is replaced by the structure
constants). Throughout this paper, we assume that none of the
denominators like $\hat k\cdot p_i$ vanish for any value of the
light like vector $\hat k$ (so that the inverses $1/\hat k\cdot p_i$ are
well defined).

\section{Transport equation for the noncommutative $U(N)$ theory}

In contrast to the $U(1)$ case, the derivation of the transport
equation, in the case of noncommutative $U(N)$ theory, presents 
a few challenges some of which we discuss below. To appreciate the
difficulties, let us recapitulate very briefly how the derivation of
the transport equation in a conventional $SU(N)$ theory is carried
out. In this case, the standard equations for the motion of a
relativistic particle carrying a color charge and interacting with an
external Yang-Mills field have the forms 
\be\label{20}
m{d x^\mu\over d\tau} = k^\mu;\;\;\;
m{d k^\mu\over d\tau} = g\, Q^a\, F^{\mu\nu\, a}k_\nu;\;\;\;
a=1,2,\cdots,N^2-1,
\ee
where $\tau$ is the proper time of the particle and $Q^a$'s are the
non-Abelian  color charges. To derive the transport equation, one
needs to supplement these with the evolution equation for the color
charges, 
\begin{equation}
m {dQ^{a}\over d\tau} = -g f^{abc} (k\cdot A^{b}) Q^{c}\label{charge}
\end{equation}
which can be obtained from a quantum field theory in a mean-field
theoretic  manner \cite{Brown:1979bv} and is consistent with the constraint
on the charges
\begin{equation}
Q^{a}Q^{a} = q_{2} ={\rm constant}\label{casimir}
\end{equation}
where $q_{2}$ represents the second Casimir of the group $SU(N)$.

Together, Eqs. (\ref{20}) and (\ref{charge}) are known as Wong's
equations \cite{wong:1970fu} and lead, 
in the commutative \sun case, to the transport equation in a
straightforward manner. In going to a noncommutative Yang-Mills
theory, first of all, we have to generalize the symmetry group to $U(N)$ which
is straightforward. Generalizing equations (\ref{20}) also poses no
particular problem. However, since in a noncommutative Yang-Mills
theory only a few representations of the gauge group are allowed
\cite{Chaichian:2001mu}, a classical color
charge is not conceptually meaningful (There are, of course, also
difficulties associated with finding a sufficiently localized state to
carry out a mean field calculation, but we will not go into
these.). Even if one formally assumes the existence of such a
classical charge, there are still difficulties associated with the
measure in the color space in the following way. It is known that it
is impossible to find local, gauge invariant observables in a
noncommutative gauge theory. Then, a constraint such as
(\ref{casimir}) is no longer gauge invariant and, therefore, is not
meaningful. As a result, the definition of the integration measure in
the color space, in such a classical theory, is unclear at the
present. Furthermore, in the presence of a constraint involving the
third  Casimir (which involves the symmetric structure constants of
the group, as would be the case in the $U(N)$ theory), the surface
terms arising from integration by parts are non-trivial and
technically much harder to calculate.

Thus, we do not fully understand a first principle derivation of the
transport equation, in the case of the noncommutative $U(N)$ gauge
theory, starting from the analog of Wong's equations and this question
is presently under study. However, motivated by the discussions in the
case of the $U(1)$ theory and taking guidance from the actual
perturbative calculations in \cite{Brandt:2002aa}, we propose a
transport equation for the noncommutative $U(N)$
theory which has the correct limiting behavior and reproduces the
perturbative results of \cite{Brandt:2002aa}. To discuss the transport equation,
let us generalize the
conventional definition of the current to this case
\be\label{21}
J_\mu^C(x) = g\sum_{{\rm }}\int dK\, dQ \, Q^C k_\mu f(x,k, Q),
\ee
where $f(x,k,Q)$ is the appropriate distribution function and we are
assuming that the measure for the integration over the color charges
is understood. Here, $C=0,1,2,\cdots , N^{2}-1$ represents the $U(N)$
indices. We use the notations and conventions in \cite{Bonora:2000ga}. Similarly,
the stress tensor for the particles can be written as
\be\label{22}
T^{\mu\nu}(x) = \sum_{{\rm }}\int dK\,dQ\, k^\mu\, k^\nu
f(x,k,Q).
\ee
It is clear now that, for these to reduce to the current and the
stress tensor studied in the last section for $N=1$, both $f$ and
$T^{\mu\nu}$ must transform under the adjoint representation of the
$U(1)$ subgroup of the $U(N)$ group. Thus, we have,
\begin{equation}
D_{\mu}f = \partial_{\mu}f - {ig\over 2} d^{BCC} (A_{\mu}^{B}\star f -
f\star A_{\mu}^{B}) = \partial_{\mu}f - {ig\over 2} d^{BCC}\,
[A_{\mu}^{B}, f]_{\rm M}\label{covariant}
\end{equation}
so that the conservation equation for the stress tensor can be written
as 
\be\label{23}
D_{\mu} T^{\mu\nu} = \partial_\mu T^{\mu\nu} - {i g\over 2} d^{BCC}
[A_\mu^{B},T^{\mu\nu}]_{\rm M} = -F^{\mu\nu\; B}\star J^B_{\mu},
\ee
Here, $d^{ABC}$ represent the completely symmetric structure constants of 
the gauge group $U(N)$ (see \cite{Bonora:2000ga} for details). 
Note that the above form of (\ref{23}) reduces to
Eq. (\ref{conservation})  when
$N=1$ and to the conventional equation in the  
commutative limit, $\theta\rightarrow 0$, when the Moyal bracket
vanishes, in which case $T^{\mu\nu}$ is manifestly gauge
invariant. The current, on the other hand, has to satisfy the standard
covariant conservation law
\begin{equation}
D_{\mu}^{AC} J^{\mu\,C} = \partial_{\mu} J^{\mu\, A} - {ig\over 2} d^{ABC}
[A_{\mu}^{B}, J^{\mu\, C}]_{\rm M} + {g\over 2}\,f^{ABC}
\{A_{\mu}^{B}, J^{\mu\, C}\}_{\rm M} = 0\label{cc}
\end{equation}
where $f^{ABC}$ represent the anti-symmetric structure constants of
$U(N)$ and  we have defined $\{A,B\}_{\rm M} = A\star B + B\star A$.

Using the above properties of the current and the stress tensor,
we propose the  following collisionless transport equation for the
noncommutative $U(N)$ theory, which generalizes
Eq. (\ref{12}).
\be\label{24}
(k\cdot {\cal D}) f(x,k,Q) = g Q^A k^\mu \, F_{\mu\nu}^A\star
{\partial f(x,k,Q)\over\partial k_\nu},
\ee
where  ${\cal D}_\mu$ is given by
\be\label{25}
{\cal D}_\mu f = \partial_\mu f + {i\, g\over 2}\, d^{ABC}\, Q^A
\left[A_\mu^B,{\partial f\over\partial Q^C}\right]_{{\rm M}} +
{g\over 2} \, f^{ABC} Q^A\left\{A_\mu^B,{\partial f\over\partial Q^{C}}
\right\}_{{\rm M}}.
\ee
We observe that the above expression reduces, in the limit 
$\theta\rightarrow 0$, to the expected classical transport equation
for the conventional non-Abelian $SU(N)$ 
Yang-Mills theory. However, as we have mentioned, our main interest here 
is concerned with the hard thermal effects  in the region $\tpte\gg 1$.

An iterative solution of the transport equation (\ref{24})
allows us to evaluate in an efficient way the hard thermal loop
contributions. For example, in order to derive the high temperature 
behavior of the two and three-gluon functions, we find recursively 
the leading order  corrections to the Bose-Einstein distribution 
$f^{(0)}$ which are given by
\be\label{26}
f^{(1)}(x,k,Q)={1\over k\cdot\partial}\, Q^B\left(k\cdot\partial
A_\mu^B - \partial_\mu\,k\cdot A^B\right){\partial f^{(0)}
\over\partial k_\mu}
\ee
\begin{eqnarray}\label{27}
f^{(2)}(x,k,Q)& = &{i\,Q^A\over 2 k\cdot\partial}\left\{-
d^{ABC}  \left[k\cdot A^B,{\partial f^{(1)}\over
\partial Q^C}\right]_{{\rm M}}
+i\, 
f^{ABC}  \left\{k\cdot A^B,{\partial f^{(1)}\over
\partial Q^C}\right\}_{{\rm M}}\right.\nnbb
&-& \left.\left(d^{ABC} \left[k\cdot A^B,A_\nu^C\right]_{{\rm M}}
              +i\,f^{ABC} \left\{k\cdot A^B,A_\nu^C\right\}_{{\rm M}}
\right){\partial f^{(0)}\over \partial k_\nu}
- 2i (k\cdot
\partial A^{\nu A} - \partial^{\nu} k\cdot A^{A})\star {\partial
f^{(1)}\over \partial k^{\nu}}\right\}
\end{eqnarray}
Once again, it can be checked that the last term in (\ref{27}) leads
to a subleading contribution at high temperature.
So, we neglect this term. 
Proceeding as in the previous case, and using the 
normalization 
\be\label{28a}
\int dQ Q^A\, Q^B = N\, \delta^{AB},
\ee 
one arrives in a straightforward manner, at the following hard thermal 
amplitudes
\be\label{28}
\Pi_{\mu\nu}^{AB}(p)  = N\, \delta^{AB}\, \Pi_{\mu\nu}(p),
\ee
where the gluon self-energy $\Pi_{\mu\nu}$ is given in Eq. (\ref{17}),
and 
\be\label{29}
\Gamma^{ABC}_{\mu\nu\lambda}(p_1,p_2,p_3) = N \left[
f^{ABC}\cos\left({p_1^\alpha\theta_{\alpha\beta}p_2^\beta\over 2}\right) +
d^{ABC}\sin\left({p_1^\alpha\theta_{\alpha\beta}p_2^\beta\over 2}\right)   
\right]\Gamma_{\mu\nu\lambda}(p_1,p_2,p_3)
\ee
where the three point function $\Gamma_{\mu\nu\lambda}(p_1,p_2,p_3)$ is
given  in Eq. (\ref{30p}). Note that the above results reduce to the 
ones in the $U(1)$ case, when we set $f^{ABC}=0$  
and $d^{ABC}$ equal to $1$.

The above hard thermal amplitudes, which are proportional to 
$T^2$, are gauge invariant and satisfy simple Ward identities. For 
example, one can easily verify the transversality property
\be\label{30}
p^\mu\,\Pi_{\mu\nu}^{AB}(p) = 0,
\ee
as well as the Ward identity which relates the two and three point gluon
functions
\be\label{31}
p_3^\lambda\,\Gamma_{\mu\nu\lambda}^{ABC}(p_1,p_2,p_3)=
i\, g\,\left[
f^{ABE}\cos\left({p_1^\alpha\theta_{\alpha\beta}p_2^\beta\over 2}\right) +
d^{ABE}\sin\left({p_1^\alpha\theta_{\alpha\beta}p_2^\beta\over 2}\right) 
\right]\left[\Pi_{\mu\nu}^{EC}(p_1) - \Pi_{\nu\mu}^{CE}(p_2)\right]
\ee

\section{The effective action for hard thermal loops}
In order to determine the effective action which generates the 
hard thermal amplitudes, we will first derive a
manifestly covariantly conserved expression for the current.
To this end, it is convenient to write the current in Eq. (\ref{21}) in a
form where the only remaining integration to be performed is that over
the angular directions of the unit vector $\hat k$, namely
\be\label{32}
J^C_{\mu}(x) = \int{d\Omega\over 4\pi}{j}_\mu^C(x, \hat k),
\ee
The quantity ${j}_\mu^C(x, \hat k)$ may be evaluated from Eq. (\ref{21}) as
\be\label{33}
{j}_\mu^C(x, \hat k) = 2\,{g\over \pi^2} \int |\vec k|^2 d|\vec k| \, d k_0
\theta(k_0)\,\delta(k^2)\, k_\mu \, \int dQ\, Q^C
f(x,k,Q),
\ee
where we have considered massless particles with two helicities.

Using the transport equation (\ref{24}), it may be verified that,
as far as the leading thermal contributions are concerned,
${j}^C_\mu(x, \hat k)$ effectively satisfies the constraint 
\be\label{34}
\left(\hat k \cdot {\cal D}\right) {j}_\mu^C(x, \hat k) =2\,
{N\,g^2\over \pi^2} \hat k_\mu\,\int 
|\vec k|^2 d|\vec k| \, d k_0 \theta(k_0)\,\delta(k^2)
\, F_{\alpha\beta}^C\, k^\alpha\,{\partial f^{(0)}\over 
\partial k_\beta},
\ee
where ${\cal D}$ is the Moyal covariant derivative given by
Eq. (\ref{25}).
This relation may be solved for ${j}^C_\mu(x, \hat k)$ and the
result substituted into Eq. (\ref{32}). Then, after performing an
integration by parts and doing the $|\vec k|$ and $k_0$ integrations,
one arrives at the following form for the current
\be\label{35}
{\bf J}_\mu = - {g^2\, T^2\over 6}\, N\, \int {d\Omega\over 4\pi}\left(
{\bf I}\; 
\eta_{\mu\beta} - \hat k_\mu {1\over \hat k\cdot {\bf  D}} 
{\bf  D}_\beta\right) 
{1\over \hat k\cdot {\bf  D}} \hat k_\alpha\, {\bf F}^{\alpha\beta},
\ee
where we have used the matrix notation:
${\bf J}_\mu = t^C\, J_\mu^C$, ${\bf A}_\mu = t^C\, A_\mu^C$,
${\bf F}_{\mu\nu} = t^C\, F_{\mu\nu}^C$ ($t^C$ are the generators of the
$U(N)$ gauge group in the fundamental representation and ${\bf I}$ is
the identity matrix). This current is manifestly covariantly conserved
\be
{\bf D}^\mu\, {\bf J}_\mu = \partial^\mu\, {\bf J}_\mu -i\,g\,
[{\bf A}^\mu,{\bf J}_\mu]_{M} = 0.
\ee
As we have seen, this property is crucial to ensure the gauge
invariance of the effective action $\Gamma\left[{\bf A}\right]$ which generates the hard
thermal amplitudes. 
From an examination of these amplitudes [see, for example, 
Eqs. (\ref{28}-\ref{31})], one notices the following properties
of the angular integrands:
\begin{itemize}
\item[(a)] The non-localities, in configuration
space,  have the form of products of operators $(k\cdot\partial)^{-1}$.
\item[(b)] They have a Lorentz covariant structure, which
  involves homogeneous functions of $k$ of degree zero.
\item[(c)] They are gauge invariant and satisfy simple Ward
identities, similar to those of the tree amplitudes.
\end{itemize}
Using analogous arguments to those given in references 
\cite{taylor:1990ia,frenkel:1991ts,brandt:1993mj}, 
one can show that these properties, together with the results for the
lowest order amplitudes, are sufficient to fix uniquely the effective
action, which is expected to be
\be\label{55}
\Gamma\left[{\bf A}\right] = {g^2\, T^2\over 6}\, N\, \int {d\Omega\over 4\pi}
\tr\,\left[
\left({\hat k^\alpha \over \hat k\cdot {\bf D}}  
{\bf F}_{\mu\alpha}\right)
\star   
\left({\hat k_\beta  \over \hat k\cdot {\bf D}}  
{\bf F}^{\mu\beta }\right)
\right].
\ee
Here, $\tr$ stands for the trace over color matrices as well as for
the appropriate space-time integrations and the operator 
$1/\hat k \cdot {\bf D}$ may be represented perturbatively as
a series of nested Moyal commutators.
The above form is manifestly gauge invariant. 
A straightforward calculation shows that
it generates the same two- and three-point functions as the one
obtained from the current.
Given the uniqueness
of the action, this is sufficient to ensure that
Eq. (\ref{55}) should represent the correct generating functional of hard
thermal loops in the noncommutative $U(N)$ theory.

The result (\ref{55}) for the effective action can also be obtained in
a more direct way. To this end,
we need to show that the current (\ref{35}) which satisfies the equation
\be
{\bf J}_\mu = t^C {\delta \Gamma\left[{\bf A}\right]\over \delta A_\mu^C}
\ee
is derivable from the action
\be\label{59b}
\Gamma\left[{\bf A}\right] = \int {d\Omega \over 4 \pi}\gamma(\hat k)\equiv
{g^2\,T^2\over 6}\,N\,\int {d\Omega \over 4 \pi}\tr\,\left[
\left({     
{\hat k^\alpha \over \hat k\cdot {\bf D}} {\bf F}_{\mu\alpha}
}\right) \star \left({
{\hat k_\beta \over \hat k\cdot {\bf D}} {\bf F}^{\mu\beta}
}\right)\right].
\ee
In order to verify the above property, we note that under a
variation  $\delta {\bf A}_\mu$ of the gauge field, one must have the
relation
\be\label{59c}
\delta \gamma(\hat k) = \tr\,\left[
{{\bf j}}^{\mu}(x,\hat k)\star
\delta {\bf A}_{\mu}(x)\right].
\ee
To check this equation, we take advantage of the fact that it is gauge
invariant, and compute each side in a gauge in which
\be\label{59d}
k^\mu\, {\bf A}_\mu = 0.
\ee
Evaluating the left and right hand side of Eq. (\ref{59c}),
respectively from the relations (\ref{59b}) and (\ref{35}),
we find that in this gauge both sides are equal to
\be\label{59e}
-{g^2\, T^2\over 3}\, N \,
\tr \,\left[\left({\bf A}^{\mu}-k^{\mu}\int_{-\infty}^0\,du\,
\partial^{\nu}{\bf A}_{\nu}(x+k\,u)\right)\star\delta {\bf A}_{\mu}\right].
\ee
Since Eq. (\ref{59c}) is gauge invariant and holds in the gauge
(\ref{59d}), it must be true in any gauge. 

Thus, Eq. (\ref{55}) gives the correct expression which describes the effective
action for hard thermal loops in the strong noncommutative regime
$\tpte \gg 1$. On the other hand, in the limit $\theta\rightarrow 0$,
the form (\ref{55}) also reduces to the expected effective action in
the commutative theory \cite{Braaten:1990it}.
This result may be considered as a first step towards the
construction of an effective action for the noncommutative Yang-Mills
theory at finite temperature, which may be of interest for a further
understanding of noncommutative QCD.
%
\begin{acknowledgments}
This work was supported in part by US DOE Grant number DE-FG
02-91ER40685, by CNPq and by FAPESP. \hbox{D. G. C. M.} 
would like to thank the  {\it Universidade de S\~ao Paulo}
for its hospitality.
\end{acknowledgments}

\end{document}